\documentclass[sigconf]{acmart}

\AtBeginDocument{%
  }

\copyrightyear{2026}
\acmYear{2026}
\setcopyright{cc}
\setcctype{by}
\acmConference[SoCC '26]{ACM Symposium on Cloud Computing V.2}{November 18--20, 2026}{Singapore, Singapore}
\acmBooktitle{ACM Symposium on Cloud Computing V.2 (SoCC '26), November 18--20, 2026, Singapore, Singapore}
\acmDOI{10.1145/3828161.3857897}
\acmISBN{979-8-4007-2816-7/2026/11}

\usepackage[nolist]{acronym} 
\usepackage{subcaption}

\newcommand{\parag}[1]{
  \par\medskip
  \noindent\textbf{#1.} \ignorespaces
}

\begin{document}

\author{Anton Altenbernd}
\orcid{0009-0004-4355-0122}
\affiliation{
  \institution{Technische Universität Berlin}
  \city{Berlin}
  \country{Germany}
}
\email{a.altenbernd@tu-berlin.de}

\author{Dylan Everingham}
\orcid{0009-0003-8847-6221}
\affiliation{
  \institution{Technische Universität Berlin}
  \city{Berlin}
  \country{Germany}
}
\email{everingham@tu-berlin.de}

\author{Philipp Wiesner}
\orcid{0000-0001-5352-7525}
\affiliation{
  \institution{University of Cambridge}
  \city{Cambridge}
  \country{United Kingdom}
}
\email{pw560@cam.ac.uk}

\author{Odej Kao}
\orcid{0000-0001-6454-6799}
\affiliation{
  \institution{Technische Universität Berlin}
  \city{Berlin}
  \country{Germany}
}
\email{odej.kao@tu-berlin.de}

\acrodef{SDC}{Silent Data Corruption}
\acrodef{SDC}{Silent Data Corruption}
\acrodef{LLM}{Large Language Model}
\acrodef{ECC}{Error-Correction Code}
\acrodef{MSB}{Most Significant Bit}
\acrodef{ABFT}{Algorithm-Based Fault Tolerance}
\acrodef{GEMM}{General Matrix-Matrix Multiplication}
\acrodef{GEMV}{General Matrix-Vector Multiplication}
\acrodef{CSNB}{Cauchy-Schwarz Norm Bounds}
\acrodef{MoE}{Mixture of Experts}
\acrodef{KV}{key--value}

\title[Revisiting Cauchy-Schwarz Norm Bounds for SDC Detection in LLM Inference]{%
Revisiting Cauchy--Schwarz Norm Bounds for SDC Detection in LLM Inference}

\begin{abstract}
Large Language Models (LLMs) execute billions of floating-point operations per output token across many hardware accelerators. At this scale, hardware-induced faults can bypass system-level detection mechanisms and cause Silent Data Corruption (SDC). Algorithm-Based Fault Tolerance (ABFT) is the state-of-the-art technique for protecting matrix multiplication from such faults, but it incurs non-negligible computational overhead and requires thresholds to distinguish hardware-induced errors from floating-point roundoff. In this paper, we revisit Cauchy--Schwarz Norm Bounds (CSNB) as a low-overhead alternative for detecting SDC during LLM inference. We compare CSNB against ABFT through controlled fault injection and evaluate the runtime overhead of both verification and threshold determination across multiple model sizes, batch sizes, and generation lengths.

Our results show that, on average, CSNB retains 99\% of the fault detection capability of ABFT for the evaluated linear layers in dense LLMs, while reducing verification overhead by 87\% and threshold determination overhead by 67\%. For Mixture-of-Experts (MoE) router gate computations, however, CSNB retains only 55\% of the fault detection capability of ABFT. These findings suggest that CSNB is a promising low-overhead alternative for protecting linear layers in dense LLMs, while checksum-based verification remains preferable for the evaluated MoE router gate.
\end{abstract}

\ccsdesc[500]{Computer systems organization~Reliability}
\ccsdesc[500]{Computing methodologies~Natural language processing}

\keywords{Large Language Models, Silent Data Corruption, Algorithm-Based Fault Tolerance, Cauchy--Schwarz Inequality, Fault Detection}

\maketitle

\section{Introduction}
\label{sec:introduction}

Since the introduction of the Transformer architecture~\cite{vaswani2017attention}, \acfp{LLM} have become a central component of modern artificial intelligence systems. Large-scale LLM inference is executed across many hardware accelerators~\cite{pope2023efficiently} and requires billions of floating-point operations to generate each output token~\cite{kaplan2020scaling}. Consequently, even low-probability hardware faults can affect inference, making reliable execution an important concern.

A major challenge arises from \ac{SDC}, which silently bypasses system-level detection mechanisms such as \acp{ECC}~\cite{papadimitriou2023silent}. 
As a result, incorrect computations may remain undetected for extended periods, compromising application correctness and complicating debugging efforts~\cite{dixit2021silent, altenbernd2026exploring, ma2025understanding}. 
\ac{SDC} can be caused by transient faults induced by external factors such as radiation~\cite{ziegler1996terrestrial, mukherjee2005soft, baumann2005radiation}, as well as by internal hardware faults resulting from manufacturing defects or component aging~\cite{dixit2021silent, hochschild2021cores, wang2023understanding}. 
Although these faults are relatively rare in conventional computing environments, their likelihood increases in large-scale systems due to the sheer number of devices, sustained workload-induced stress, and elevated operating temperatures~\cite{dixit2022detecting, wang2023understanding}. 
As a result, ensuring resilience against \ac{SDC} is essential for reliable LLM inference. 
Emerging proposals for space-based data centers~\cite{starcloud2025} further motivate research in this area.

Transformer-based inference is largely composed of linear operations, including the query, key, value, and output projections in the attention mechanism, the gate, up, and down projections in the feed-forward network, the \ac{MoE} routing, and the language modeling head~\cite{vaswani2017attention, shazeer2017outrageously}. 
These operations are implemented as matrix multiplications, account for the majority of the computational workload~\cite{he2025waferllm}, and, due to their linearity, satisfy algebraic invariants that make them a natural target for fault-tolerance mechanisms.

\parag{An alternative to ABFT}
The state-of-the-art technique for protecting matrix multiplications underlying linear layers is \ac{ABFT}. 
It exploits the algebraic structure of matrix multiplication by augmenting the input operands with checksums, providing error detection with significantly lower overhead than redundant execution~\cite{huang1984algorithm}.
However, \ac{ABFT} still introduces non-negligible computational overhead and requires threshold determination to compensate for floating-point roundoff errors, which can result in loose verification bounds and additional overhead~\cite{gao2026v}.
This raises the question of whether simpler verification mechanisms can provide comparable fault detection while reducing runtime overhead for LLM inference.

An alternative approach exploits the Cauchy--Schwarz inequality~\cite{strang2000linear} to derive an upper bound on the magnitude of linear layer outputs that can be checked during execution.
Despite its conceptual simplicity and low computational cost, the resulting bound is tight only when the corresponding vectors are linearly dependent, which can make it overly conservative for error detection. 
However, to the best of our knowledge, the potential of \ac{CSNB} for \ac{SDC} detection during \ac{LLM} inference has not been investigated.

\parag{Contributions}
In this paper, we revisit \ac{CSNB} as a low-overhead alternative for detecting \ac{SDC} during \ac{LLM} inference and, to the best of our knowledge, provide the first systematic evaluation of norm-bound-based verification in this setting. We make the following contributions:
\begin{itemize}
    \item We compare \ac{CSNB} against V-ABFT~\cite{gao2026v}, a recent \ac{ABFT} extension with online threshold determination, through controlled fault injection across four dense \acp{LLM} and a representative \ac{MoE} model, spanning both prefill and decoding and fault multiplicities from 1 to 256 bit flips.
    \item We quantify the runtime overhead of both the verification stage and threshold determination. The verification stage is implemented directly within fused Triton-based~\cite{tillet2019triton} matrix multiplication kernels, whereas threshold determination is evaluated separately using a Python implementation across varying model sizes, batch sizes, and generation lengths.
    \item We show that \ac{CSNB} achieves comparable fault detection capability to V-ABFT on the evaluated linear layers in dense \acp{LLM} at substantially lower overhead. Checksum-based verification remains preferable for the evaluated \ac{MoE} router gate computations.
\end{itemize}

\section{Background}
\label{sec:background}

This section introduces the two verification mechanisms compared throughout the remainder of the paper.

\parag{Algorithm-Based Fault Tolerance (ABFT)} In the classical \ac{ABFT} scheme for matrix multiplication~\cite{huang1984algorithm}, the objective is to protect the computation \begin{equation}C = AB,\end{equation} where \(A \in \mathbb{R}^{m \times k}\), \(B \in \mathbb{R}^{k \times n}\), and \(C \in \mathbb{R}^{m \times n}\). The key idea is to augment the input matrices with checksum rows and/or columns before computation. These checksums are propagated together with the original data during the matrix multiplication, allowing the correctness of the result to be verified without recomputing the entire operation. For row checksums, the augmented computation is given by \begin{equation}\begin{bmatrix}A \\(\mathbf{1}^\top A)\end{bmatrix}B=\begin{bmatrix}C \\\mathbf{1}^\top C\end{bmatrix},\end{equation} where the last row represents the propagated checksum of the output matrix. Verification is performed by comparing this propagated checksum with the checksum recomputed from the computed result, \begin{equation}R = \mathbf{1}^\top C - (\mathbf{1}^\top A)B.\end{equation}

In exact arithmetic, where \(C = AB\), the residual satisfies \(R = 0\). In floating-point arithmetic, however, rounding errors accumulate during both the matrix multiplication and checksum propagation, causing \(R\) to be nonzero even in the absence of hardware faults. Consequently, practical \ac{ABFT} implementations verify the result by checking \begin{equation}\|R\| \leq \tau,\end{equation} where \(\tau\) is a verification threshold chosen to distinguish floating-point roundoff from \ac{SDC}~\cite{braun2014abft,gao2026v}. The effectiveness of ABFT therefore depends critically on selecting a threshold that is sufficiently large to avoid false positives while remaining small enough to detect faults reliably. 

V-ABFT~\cite{gao2026v} addresses threshold selection by directly modeling the \ac{ABFT} verification difference caused by floating-point rounding. It derives a variance-based adaptive threshold using row-wise statistics such as the mean, minimum, and maximum. The authors report that this yields tighter thresholds than A-ABFT~\cite{braun2014abft}, an earlier analytical threshold determination method, while maintaining a zero false-positive rate in the evaluated configurations. In this work, we compare against V-ABFT with respect to fault detection capability and runtime overhead of threshold determination, as it is the most recent floating-point-aware \ac{ABFT} method and provides a tighter threshold determination than A-ABFT.

\parag{Cauchy--Schwarz Norm Bounds (CSNB)} An alternative approach exploits the Cauchy--Schwarz inequality~\cite{strang2000linear} to derive bounds on the output entries of the matrix multiplication. The Cauchy--Schwarz inequality provides an upper bound on the magnitude of an inner product. For vectors \(x,y \in \mathbb{R}^{k}\), it states that \begin{equation}|x^\top y| \leq \|x\|_2 \|y\|_2,\end{equation} with equality if and only if \(x\) and \(y\) are linearly dependent. Since each entry of a matrix product is an inner product between a row of \(A\) and a column of \(B\), the inequality immediately yields \begin{equation}|C_{ij}| = |a_i^\top b_j|\leq\|a_i\|_2 \|b_j\|_2,\end{equation} where \(a_i^\top\) denotes the \(i\)-th row of \(A\) and \(b_j\) denotes the \(j\)-th column of \(B\). Applying this bound to every output entry yields an \(m \times n\) matrix of bounds, requiring the storage of all row and column norms. To reduce the storage overhead, a practical alternative is to use a single global bound based on the largest row and column norms, \begin{equation}|C_{ij}| \leq\max_{1 \leq i \leq m}\|a_i\|_2 \;\max_{1 \leq j \leq n}\|b_j\|_2,\qquad\forall\, i,j.\end{equation} 

In exact arithmetic, every output entry satisfies the Cauchy--Schwarz bound by construction. In floating-point arithmetic, both the computed matrix entries and the corresponding norms are affected by rounding errors. Unlike \ac{ABFT}, which verifies an equality, \ac{CSNB} verifies a conservative inequality. The bound is tight only when the corresponding vectors are linearly dependent and becomes even more conservative when using the global bound. In this work, we therefore evaluate outputs directly against the bound without deriving an explicit floating-point verification threshold. The resulting false-positive behavior is evaluated experimentally.

\section{Methodology}

We first describe the fault injection methodology before introducing the detection and runtime evaluation experiments.

\parag{Experimental setup}
For every experiment, the detection mechanism is incorporated by replacing selected linear layer modules with detection-enabled implementations while preserving the original model parameters and inference procedure. Inference is performed in \texttt{bfloat16} precision. All experiments are conducted on NVIDIA L40S GPUs, using a single GPU for the dense models and two GPUs for the \ac{MoE} model.

Evaluation is performed using 1{,}000 prompts from each of the GSM8K~\cite{cobbe2021training} and SQuAD v2~\cite{rajpurkar2018know} datasets. GSM8K provides mathematical reasoning tasks, whereas SQuAD v2 consists of extractive question-answering examples. The prompts are sampled uniformly at random, and for each prompt, faults are injected during either the prefill or autoregressive decoding phase of inference and for fault multiplicities of 1, 2, 64, and 256 bit flips, resulting in 8,000 fault injection runs per model, dataset, and method.

Faults are injected into the output activations immediately after the linear layer computation. If the fault is injected during decoding, the decoding step is sampled uniformly at random. A linear layer is sampled uniformly from the protected layers, followed by a random output tile (64$\times$64 elements) and a randomly selected bit position of the \texttt{bfloat16} representation. Depending on the evaluated fault multiplicity, 1, 2, 64, or 256 output activations are selected uniformly at random within the chosen output tile, and the selected bit position is flipped in each selected activation. This tile-based injection approximates the spatial locality of hardware-induced corruptions, while the affected activations and bit position are selected randomly as hardware-specific fault patterns are architecture-dependent.

\parag{Detection experiments}
For every injection, the generated token sequence is compared against a fault-free reference execution to determine whether the injected fault is masked or results in an \ac{SDC}. We therefore evaluate whether the generated sequence changes, rather than the semantic correctness of the model output. Detection is evaluated independently of whether the generated sequence changes. Consequently, every fault injection falls into one of four categories: masked and undetected, masked and detected, \ac{SDC} and undetected, or \ac{SDC} and detected. The corresponding fault-free reference executions are used to assess false-positive behavior.

The first experiment evaluates the \ac{SDC} detection capability of \ac{CSNB} on dense transformer architectures. 
We compare \ac{CSNB} against V-ABFT using four \acp{LLM} representing different transformer families: Qwen2.5-7B~\cite{qwen2024qwen205}, Mistral-7B~\cite{jiang2023mistral}, Falcon3-7B~\cite{Falcon3}, and Llama3-8B~\cite{grattafiori2024llama}. 
For each model, every linear layer is replaced by its corresponding detection-enabled implementation. Matrix multiplications performed within FlashAttention~\cite{dao2022flashattention} are excluded as kernel fusion with the attention computation prevents the direct application of fault detection schemes.

The second experiment evaluates the \ac{SDC} detection capability of \ac{CSNB} for router gate computations in \ac{MoE} architectures. 
We compare \ac{CSNB} against V-ABFT using Mixtral-8x7B~\cite{jiang2024mixtral} as a representative \ac{MoE} model. 
Unlike the first experiment, only the linear layers implementing the router gate are replaced by their corresponding detection-enabled implementations.
This isolates the effectiveness of the proposed detection mechanism for routing decisions, which determine the expert assignment for each token and therefore constitute a critical point in the inference path.

\parag{Runtime evaluation}
We separately evaluate the runtime overhead of verification and threshold determination. The runtime evaluation uses Qwen2.5-3B, Qwen2.5-7B, and Qwen2.5-14B~\cite{qwen2024qwen205} with a fixed input prompt of 256 tokens. Preliminary runtime measurements across the evaluated dense transformer architectures showed no meaningful architectural differences. Therefore, the runtime evaluation focuses on the Qwen2.5 family and studies the impact of model size, batch size, and generation length. We do not perform a corresponding runtime evaluation for the \ac{MoE} model because fault detection is applied only to the router gate computations, which represent only a small fraction of the overall inference workload. End-of-sequence token generation is disabled to ensure that every inference run generates the configured number of output tokens. Batch sizes of 1, 2, 4, 8, 16, 32, and 64 and generation lengths of 1, 64, and 256 new tokens are considered. For every model, implementation variant, batch size, and generation length, five warm-up inference runs are performed and excluded from the reported results. Subsequently, ten inference runs are measured and averaged.

The verification stage is implemented within fused Triton~\cite{tillet2019triton} matrix multiplication kernels. We compare the baseline implementation without \ac{SDC} detection against \ac{ABFT} and \ac{CSNB}. Verification overhead is reported relative to the baseline implementation without \ac{SDC} detection.

Threshold determination is evaluated independently using a Python implementation. Since no optimized implementation of threshold determination is publicly available for V-ABFT, and its implementation overlaps threshold determination with the matrix multiplication, we evaluate both V-ABFT and \ac{CSNB} independently of the matrix multiplication. This avoids confounding the comparison with implementation-specific optimizations, which could similarly be applied to \ac{CSNB}, and instead compares the computational cost of the underlying threshold determination algorithms. We compare V-ABFT and \ac{CSNB}, as well as variants that precompute quantities derived from the model parameters during initialization, namely parameter norms for \ac{CSNB} and the parameter statistics used to derive the V-ABFT thresholds. 

\begin{figure*}[!t]
    \centering
    \begin{subfigure}[t]{0.485\textwidth}
        \centering
        \includegraphics[width=\linewidth]{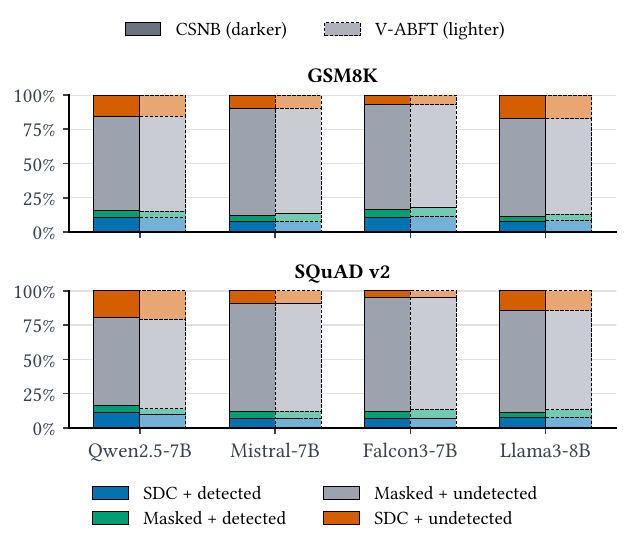}
        \caption{Fault injection with KV caching enabled.}
        \label{fig:detection_dense}
    \end{subfigure}
    \hfill
    \begin{subfigure}[t]{0.485\textwidth}
        \centering
        \includegraphics[width=\linewidth]{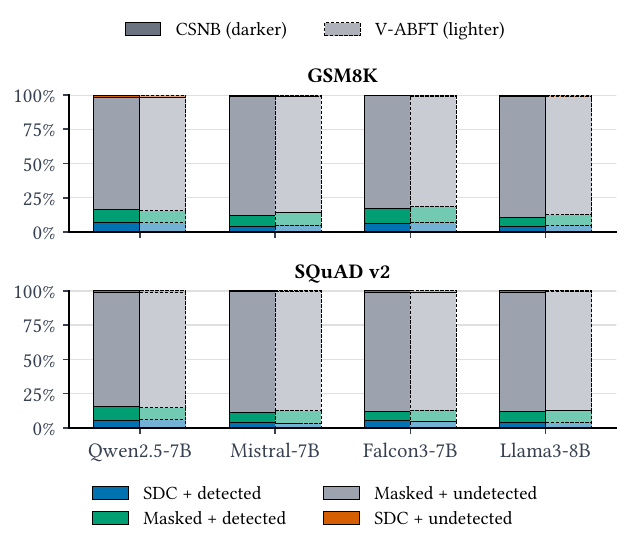}
        \caption{Fault injection with KV caching disabled.}
        \label{fig:detection_dense_no_cache}
    \end{subfigure}
    \caption{Fault detection outcomes by model, dataset, and verification method. Stacked bars show the fraction of fault injection runs in each outcome category. Darker bars with solid outlines denote \ac{CSNB}, whereas lighter bars with dashed outlines denote V-\ac{ABFT}. Across the evaluated dense transformer models, \ac{CSNB} achieves fault detection capability comparable to V-\ac{ABFT} with and without KV caching. Disabling KV caching prevents corrupted key--value activations from propagating across decoding steps, thereby reducing the fraction of undetected \ac{SDC}.}
    \label{fig:detection_dense_comparison}
\end{figure*}

\section{Evaluation}

This section shows that \ac{CSNB} achieves comparable fault detection to V-ABFT at lower runtime overhead for dense \acp{LLM}, but is less effective for \ac{MoE} router gates.

\parag{Comparable detection in dense models}
Figure~\ref{fig:detection_dense} compares the overall fault detection outcomes of \ac{CSNB} and V-ABFT across four transformer-based dense \acp{LLM} on GSM8K and SQuAD v2. 
Across both datasets and all evaluated models, the two approaches exhibit highly similar detection behavior. The majority of injected faults are masked by the network, while the fraction of undetected \ac{SDC} remains nearly identical for both methods. 
These results indicate that replacing checksum-based verification with \ac{CSNB} does not noticeably reduce fault detection capability for dense transformer inference. In addition, neither \ac{CSNB} nor V-ABFT produced false-positive detections in any fault-free execution.

\parag{KV caching enables fault propagation}
Although both \ac{CSNB} and V-ABFT detect most large perturbations, the fault injection experiments still exhibit a fraction of undetected \acp{SDC}, including faults affecting low-order mantissa bits. We therefore investigate whether KV cache propagation contributes to the observed \acp{SDC}. 
Figure~\ref{fig:detection_dense_no_cache} repeats the experiment with KV caching disabled. With KV caching enabled, corrupted key--value activations are stored in the cache and reused during subsequent decoding steps, allowing faults to propagate during autoregressive generation.
Disabling KV caching removes this propagation path by recomputing the key--value activations from scratch at every decoding step. 
As a result, the fraction of \ac{SDC} decreases, while the relative behavior of \ac{CSNB} and V-ABFT remains nearly unchanged, indicating that both approaches respond similarly regardless of whether faults propagate through the KV cache.

\begin{figure*}[!t]
    \centering
    \begin{subfigure}[t]{0.485\textwidth}
        \centering
        \includegraphics[width=\linewidth]{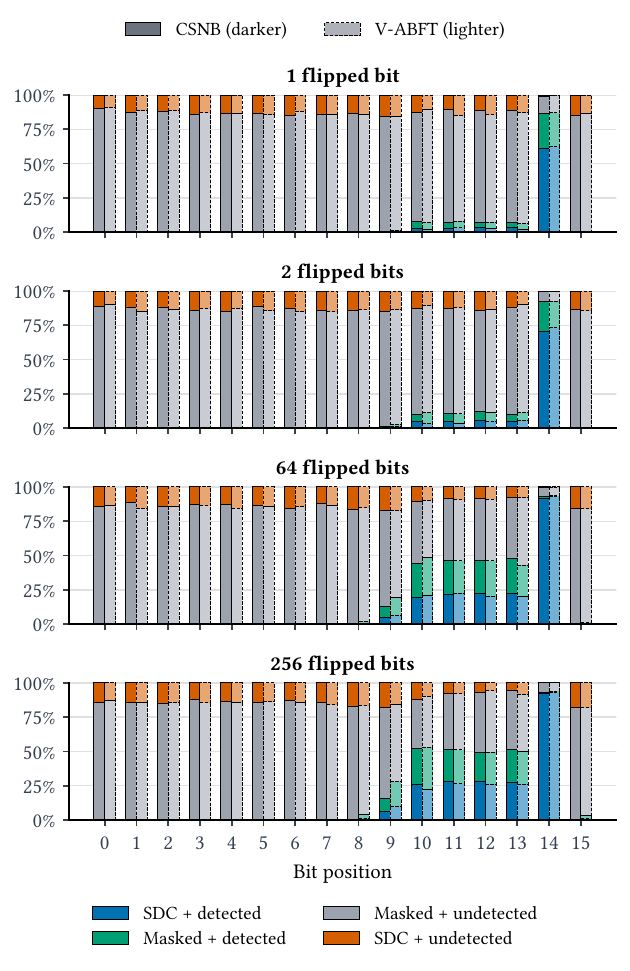}
        \caption{Dense transformer models.}
        \label{fig:detection_bits_dense}
    \end{subfigure}
    \hfill
    \begin{subfigure}[t]{0.485\textwidth}
        \centering
        \includegraphics[width=\linewidth]{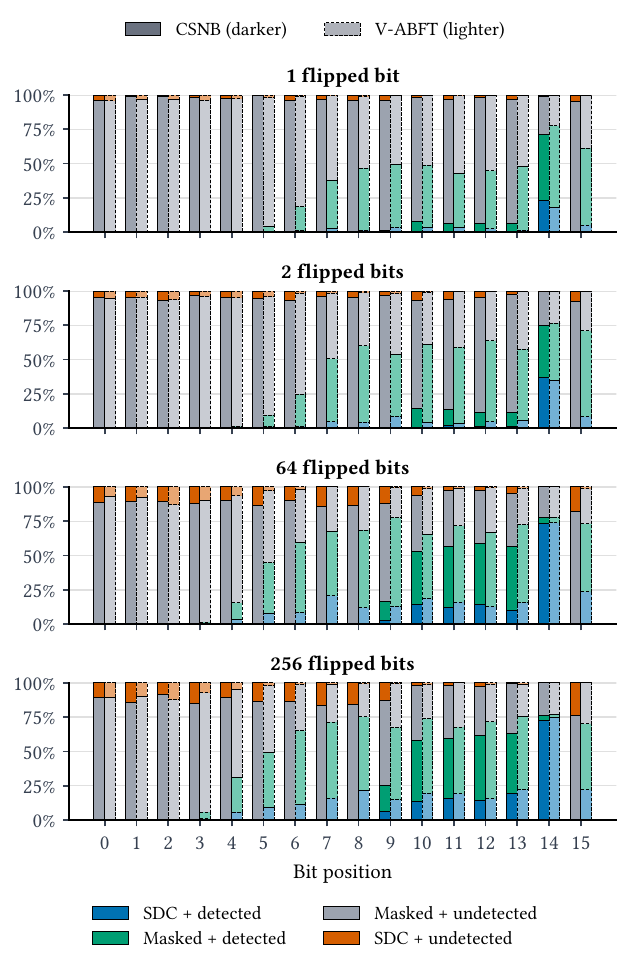}
        \caption{\ac{MoE} router gate computations.}
        \label{fig:detection_bits_moe}
    \end{subfigure}
    \caption{Fault injection outcomes by flipped bit position, with stacked bars showing the fraction of runs in each category. Darker bars with solid outlines denote \ac{CSNB}, whereas lighter bars with dashed outlines denote V-\ac{ABFT}. For both dense transformer models and \ac{MoE} router gate computations, fault detection capability increases with the number of injected bit flips. While \ac{CSNB} and V-\ac{ABFT} exhibit nearly identical fault detection behavior for dense transformer models, V-\ac{ABFT} achieves higher fault detection capability for \ac{MoE} router gate computations, particularly for faults affecting mantissa and lower exponent bits.}
    \label{fig:detection_bits}
\end{figure*}

\parag{Detection depends on location and multiplicity}
Figures~\ref{fig:detection_bits_dense} and~\ref{fig:detection_bits_moe} provide a more detailed analysis by reporting detection outcomes for each flipped bit position and fault multiplicity. 
For dense transformer models (Figure~\ref{fig:detection_bits_dense}), both approaches exhibit almost identical behavior across all evaluated bit positions. 
As expected, faults affecting the exponent bits produce a substantially higher fraction of detected \ac{SDC} than faults confined to the mantissa, while increasing the number of injected bit flips increases the overall detection rate for both methods.

In contrast, the behavior differs for \ac{MoE} router gate computations (Figure~\ref{fig:detection_bits_moe}). 
While both approaches detect faults affecting the most significant exponent bits with high probability, V-ABFT also detects a noticeable fraction of faults affecting the mantissa and remaining exponent bits, whereas \ac{CSNB} detects only a small fraction of such corruptions. 
Since router outputs directly determine expert selection, comparatively small numerical perturbations may alter routing decisions, making these computations more sensitive to verification accuracy. Consequently, \ac{CSNB} exhibits lower fault detection capability than V-ABFT for router gate computations.

\begin{figure*}[!t]
    \centering
    \begin{subfigure}[t]{0.485\textwidth}
        \centering
        \includegraphics[width=\linewidth]{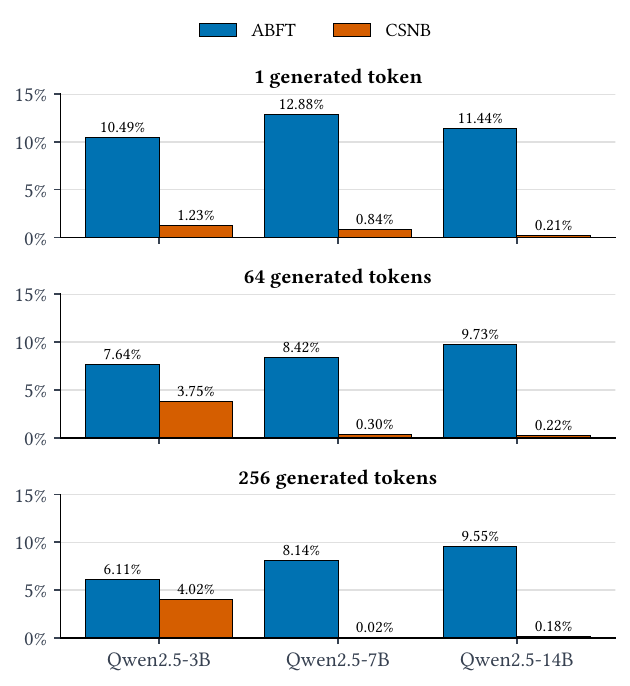}
        \caption{Verification overhead.}
        \label{fig:runtime_verification}
    \end{subfigure}
    \hfill
    \begin{subfigure}[t]{0.485\textwidth}
        \centering
        \includegraphics[width=\linewidth]{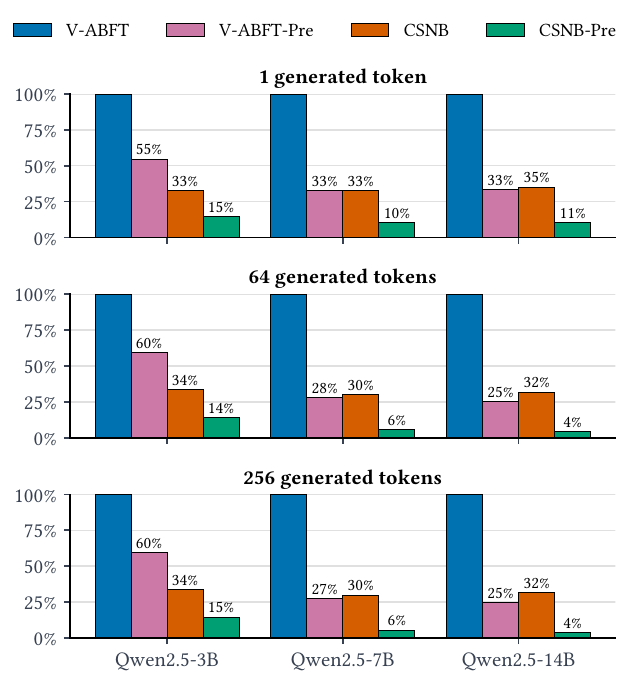}
        \caption{Threshold determination overhead.}
        \label{fig:runtime_threshold}
    \end{subfigure}
    \caption{Runtime overhead of \ac{CSNB} and \ac{ABFT}. Reported runtimes are averaged across all evaluated batch sizes. Left: verification overhead relative to the baseline implementation without \ac{SDC} detection. Right: threshold determination normalized to V-\ac{ABFT} (100\%). The precomputed (-Pre) variants cache model-dependent statistics during initialization. \ac{CSNB} consistently incurs lower verification overhead than \ac{ABFT}, while threshold determination requires substantially less computation than V-\ac{ABFT} and is further accelerated by precomputing model-dependent statistics during initialization.}
    \label{fig:runtime_results}
\end{figure*}

\parag{CSNB reduces runtime overhead}
Figure~\ref{fig:runtime_results} presents runtime results averaged across batch sizes to highlight the impact of model size and generation length. While threshold determination depends little on batch size, the verification overhead of \ac{ABFT} approaches that of \ac{CSNB} at the largest batch size (64).

Figure~\ref{fig:runtime_verification} reports the verification overhead relative to the baseline without \ac{SDC} detection.Across all evaluated configurations, \ac{CSNB} consistently incurs substantially lower verification overhead than \ac{ABFT}. While \ac{ABFT} introduces approximately 6--13\% overhead depending on model size and generation length, \ac{CSNB} remains below 5\% and approaches near zero for the larger Qwen2.5-7B and Qwen2.5-14B \acp{LLM}. This reduction is expected because \ac{CSNB} only evaluates a threshold during verification, whereas \ac{ABFT} additionally computes and compares propagated checksums.

Figure~\ref{fig:runtime_threshold} presents the runtime of threshold determination, normalized to V-ABFT. Both V-ABFT and \ac{CSNB} benefit from precomputing model-dependent quantities during initialization, reducing threshold determination time by 40--75\% and 55--87\%, respectively, compared to their online implementations. Even without precomputation, \ac{CSNB} requires only about one-third of the runtime of V-ABFT, while the precomputed \ac{CSNB} variant reduces this cost to approximately 4--15\%.

\section{Discussion}

This section discusses the implications and limitations of \ac{CSNB} and V-\ac{ABFT} for different linear layer computations.

\parag{CSNB as a practical alternative}
For the evaluated linear layers in dense \acp{LLM}, \ac{CSNB} achieves fault detection performance comparable to V-ABFT while requiring less computation during both runtime verification and threshold determination. This provides an attractive trade-off for deployment in large-scale inference systems, where every generated token requires many linear layer evaluations and even small per-layer overheads accumulate over long decoding sequences. Despite the conservative upper bound provided by the Cauchy--Schwarz inequality, one possible explanation for the comparable detection performance is that the adaptive thresholds of V-ABFT are likewise conservative for these layers. Since these thresholds are derived from statistical properties of the inputs, they may become loose depending on input distributions. Consequently, both approaches may employ verification bounds of similar practical tightness.

\parag{Verification depends on the protected computation}
The reduced effectiveness of \ac{CSNB} for \ac{MoE} router gate computations highlights a limitation of norm-bound-based verification.
Unlike the evaluated dense linear layers, the router computes logits whose relative ordering determines expert selection.
Consequently, comparatively small perturbations may alter routing decisions even when the resulting activations remain within the corresponding verification bound.
One possible explanation for the difference in detection capability lies in the verification thresholds.
The adaptive threshold of V-\ac{ABFT} depends on the number of output columns $n$ and the statistical properties of the input matrices~\cite{gao2026v}.
The evaluated Mixtral-8x7B router gate has only $n=8$ output columns, whereas the evaluated dense linear layers have substantially larger output dimensions.
This may result in a tighter V-\ac{ABFT} threshold for the router gate, while the global \ac{CSNB} bound depends on the maximum row and column norms and does not directly tighten as $n$ decreases.
Consequently, perturbations below the \ac{CSNB} verification bound may still exceed the V-\ac{ABFT} threshold.
These findings suggest that the suitability of a verification mechanism depends on the protected computation.
While \ac{CSNB} achieves comparable detection capability to V-\ac{ABFT} for the evaluated linear layers in dense  \acp{LLM}, checksum-based verification remains preferable for the evaluated \ac{MoE} router gate.
Whether this limitation extends to other \ac{MoE} router gates with more experts or to other linear layers within \ac{MoE} models has not been evaluated in this work.

\parag{Selective protection}
The presented evaluation assumes that every evaluated linear layer is protected. In practice, however, this may not be necessary. Hardware fault conditions may persist across multiple kernel executions rather than affecting only a single linear layer computation. Consequently, protecting only a subset of strategically selected layers may still provide a high probability of detecting faults while further reducing runtime overhead.

\parag{Limitations of local verification}
Beyond the differences between \ac{CSNB} and V-ABFT, the presented results also highlight a limitation shared by both approaches. Both techniques verify individual linear layer computations and therefore provide only a local view of correctness. Faults that remain below the local verification criterion may nevertheless propagate through subsequent transformer layers and eventually affect the generated output. Similarly, once corrupted key or value activations have been written to the KV cache, subsequent decoding steps reuse the corrupted values, as neither method detects errors beyond the protected linear layer. Addressing such error propagation will likely require detection mechanisms that extend beyond single linear layer computations.

\section{Conclusion}

This paper revisited \ac{CSNB} for detecting \ac{SDC} during \ac{LLM} inference.
Through controlled fault injection and runtime evaluation, we compared \ac{CSNB} against V-ABFT across dense transformer models and \ac{MoE} routing.
Our results show that, on average, \ac{CSNB} retains 99\% of the fault detection capability of V-ABFT for the evaluated linear layers in dense \acp{LLM}, while reducing verification overhead by 87\% and threshold-determination overhead by 67\%.
For the evaluated router gate computations, however, \ac{CSNB} retains only 55\% of the fault detection capability of V-ABFT.
These findings suggest that \ac{CSNB} is a promising low-overhead alternative for protecting dense linear layers in \ac{LLM} inference, while checksum-based verification remains preferable for the evaluated router gate.

\bibliographystyle{ACM-Reference-Format}
\bibliography{refs}

\end{document}